 \documentclass[useAMS,usenatbib]{mn2e}

\usepackage{graphicx}

\title[ The Jets and Disc of SS 433]
  {The Jets and Disc of SS 433 at Super-Eddington Luminosities}
\author[T. Okuda, G. V. Lipunova, and D. Molteni]{T. Okuda$^{1}$
\thanks{E-mail:okuda@hak.hokkyodai.ac.jp}, G. V. Lipunova$^{2}$,
  D. Molteni$^{3}$ \\
$^{1}$Hakodate Campus, Hokkaido University of Education, Hachiman-cho 1-2,
 Hakodate 040-8567, Japan\\
$^{2}$Sternberg Astronomical Institute, Universitetskiy pr. 13, Moscow, 119992, Russia\\
$^{3}$Dipartimento di Fisica e Tecnologie Relative, Universita di Palermo,
 Viale delle Scienza,  Palermo, 90128, Italy}
 
\begin{document}

\date{Accepted }

\pagerange{\pageref{firstpage}--\pageref{lastpage}} \pubyear{2004}

\maketitle

\label{firstpage}

\begin{abstract}
 We  examine the  jets and the disc of SS 433 at super-Eddington luminosities 
 with $\dot M \sim 600 \dot M_{\rm c}$
 by time-dependent two-dimensional radiation hydrodynamical calculations, 
 assuming $\alpha$-model for the viscosity. One-dimensional supercritical
 accretion disc models with mass loss or advection are used as the initial 
 configurations of the disc.
 As a result, from the initial advective disc models with $\alpha$ =0.001 and
 0.1, we obtain the total luminosities $\sim 2.5 \times 10^{40}$ and
$ 2.0 \times 10^{40}$ erg s$^{-1}$. 
 The total mass-outflow rates are $\sim 4\times 10^{-5}$
 and $10^{-4} {\rm M_{\odot}}$ yr$^{-1}$ and the rates of the relativistic
 axial outflows in a small half opening angle of $\sim 1^\circ$ are about 
 $10^{-6} {\rm M_{\odot}}$ yr$^{-1}$: the values generally
 consistent with the corresponding observed rates of the wind and the jets,
 respectively.
 From the initial models with mass loss but without advection, we obtain 
 the total mass-outflow and axial outflow rates smaller than or 
 comparable to the observed rates of the wind and the jets respectively,
 depending on $\alpha$. 
 In the advective disc model with $\alpha=0.1$, the initially 
 radiation-pressure dominant, optically thick disc evolves to the gas-pressure
 dominated, optically thin state in the inner region of the disc, and the inner
  disc is unstable. 
 Consequently, we find remarkable modulations of the disc luminosity and 
 the accretion rate through the inner edge. 
 These modulations manifest themselves as the recurrent hot blobs with high 
 temperatures and low densities at the disc plane, 
 which develop outward and upward and produce the QPOs-like variability
 of the total luminosity with an amplitude of a factor of $\sim$ 2 and
 quasi-periods of $\sim$ 10 -- 25 s.
 This may explain the massive jet ejection and the QPOs phenomena observed
 in SS 433. 
\end{abstract}


\section{Introduction}
 Disc accretion is an essential process for such phenomena as energetic
 X-ray sources, active galactic nuclei, and protostars.
 Since the early works by \citet{b34} and \citet{b35}, a great number 
  of papers have been devoted to studies of the disc accretion onto 
 gravitating objects. 
 When the accretion rate is not too high, the accretion disc luminosity
 is directly in proportion to the accretion rate and can be successfully
 described by the Shakura-Sunyaev (S-S) model.
 However, for the supercritical accretion rate, matter flows out of the disc,
 and the rate of accretion onto the central black hole is reduced,
 regulating the luminosity to the Eddington limit.
 The galactic microquasar SS 433 is a promising super-critically 
 accreting sterllar-mass black hole candidate and  has stimulated 
 numerous studies, because it displays remarkable observational features,
 such as its extremely high energy,
 two oppositely directed relativistic jets, and the precessing motion 
 of the jets.
 Although a number of observational and theoretical studies on SS 433 have
 been published (for detailed reviews see Margon 1984; Fabrika 2004),
 there are still many problems of the jets and the disc to be solved.
 The super-Eddington accretion discs are generally expected to
 possess vortex funnels and radiation-pressure driven jets from
 geometrically thick discs \citep{b35,b19,b12,b3}.
 Such supercritical disc models have been numerically examined 
 by two-dimensional radiation hydrodynamical calculations 
 \citep{b6,b7,b28,b30,b27,b25,b26}, especially focussing on the 
 puzzling X-ray source SS 433 \citep{b6,b7,b28,b30}.
 However, these works leave something to be desired as far as SS 433
 is concerned, because the accretion rates adopted in these studies
 are very small, compared with those estimated for SS 433.
 In this paper, we examine the properties of the jets and the disc of
 SS 433 with the plausible supercritical accretion rate. We perform
 time-dependent two-dimensional radiation hydrodynamical calculations,
 using 1D models of
 the supercritical accretion disc models with mass loss or advection 
 \citep{b20} as initial disc configurations.

\begin{figure}
\begin{center}
\includegraphics[width=8cm, height=6cm]{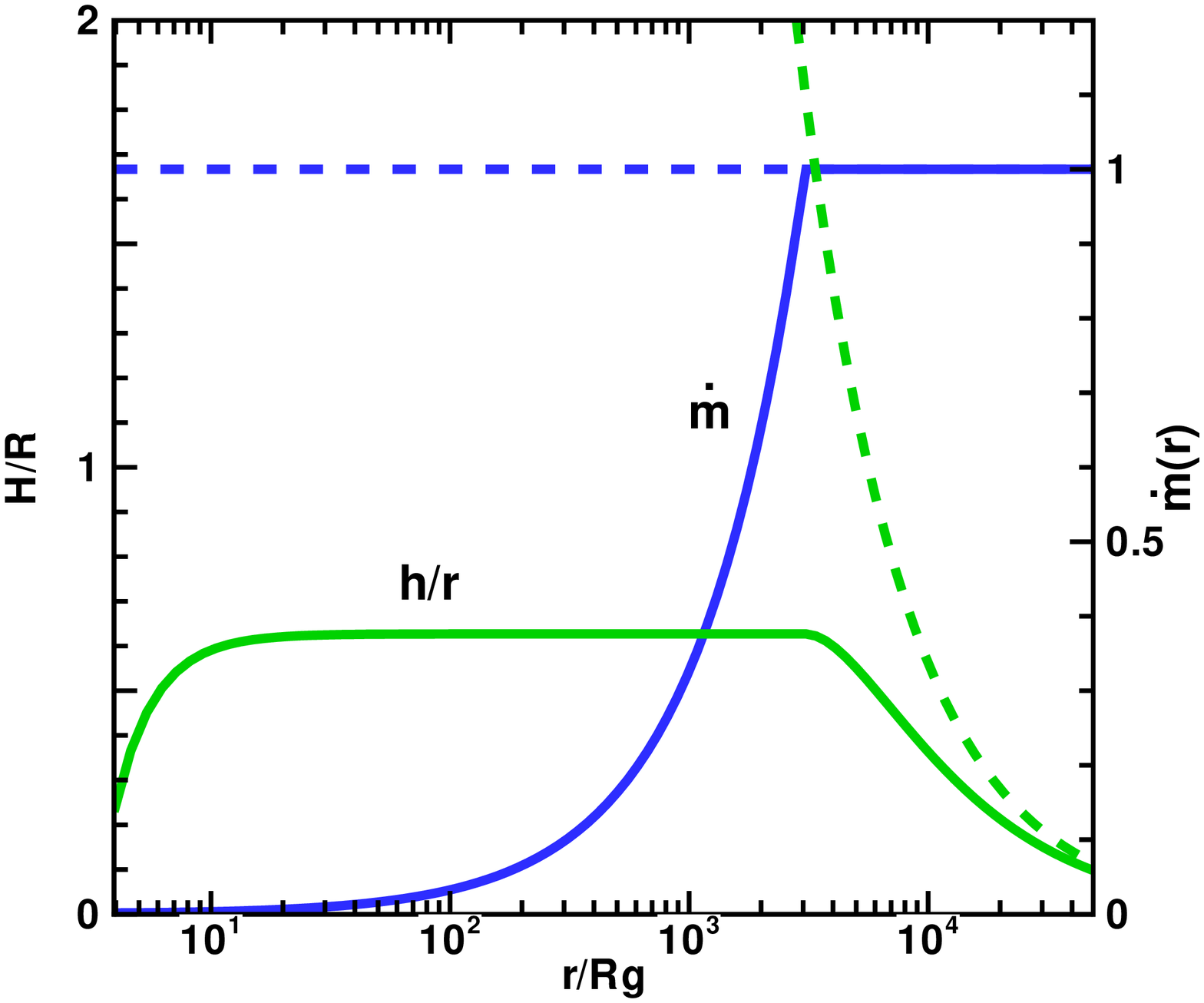}
\end{center}
\caption{Solid lines: accretion rate $\dot m(r)$ normalized to the input 
 accretion rate and relative thickness $h/r$ of the supercritical disc  with
 mass loss for the case of the input accretion rate $\dot m_0 \sim 600$ and 
 the viscosity parameter $\alpha=0.1$. Dashed lines show the solution for
 S-S model without mass loss.
   }
\label{fig:fig1}
\end{figure}

\begin{figure}
\begin{center}
\includegraphics[width=8cm, height=6cm]{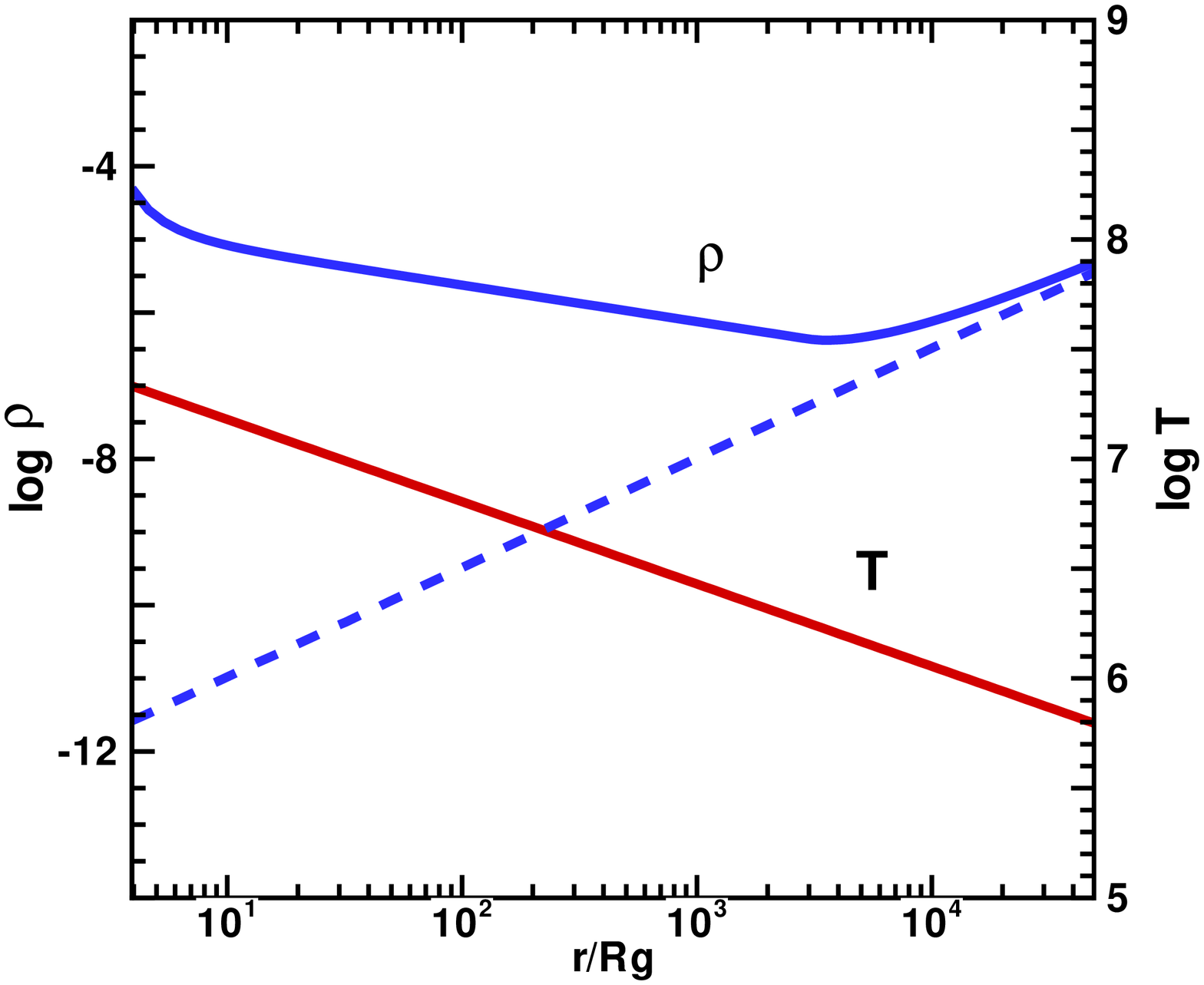}
\end{center}
\caption{Same as Fig.1 but for the profiles of temperature $T$ (K) and density
 $\rho$ (g$\;{\rm cm}^{-3}$). 
 The temperature profile is the same for the both models, as far as the 
 radiation-pressure dominant disc is concerned.
   }
\label{fig:fig2}
\end{figure}

\begin{figure}
\begin{center}
\includegraphics[width=8cm, height=6cm]{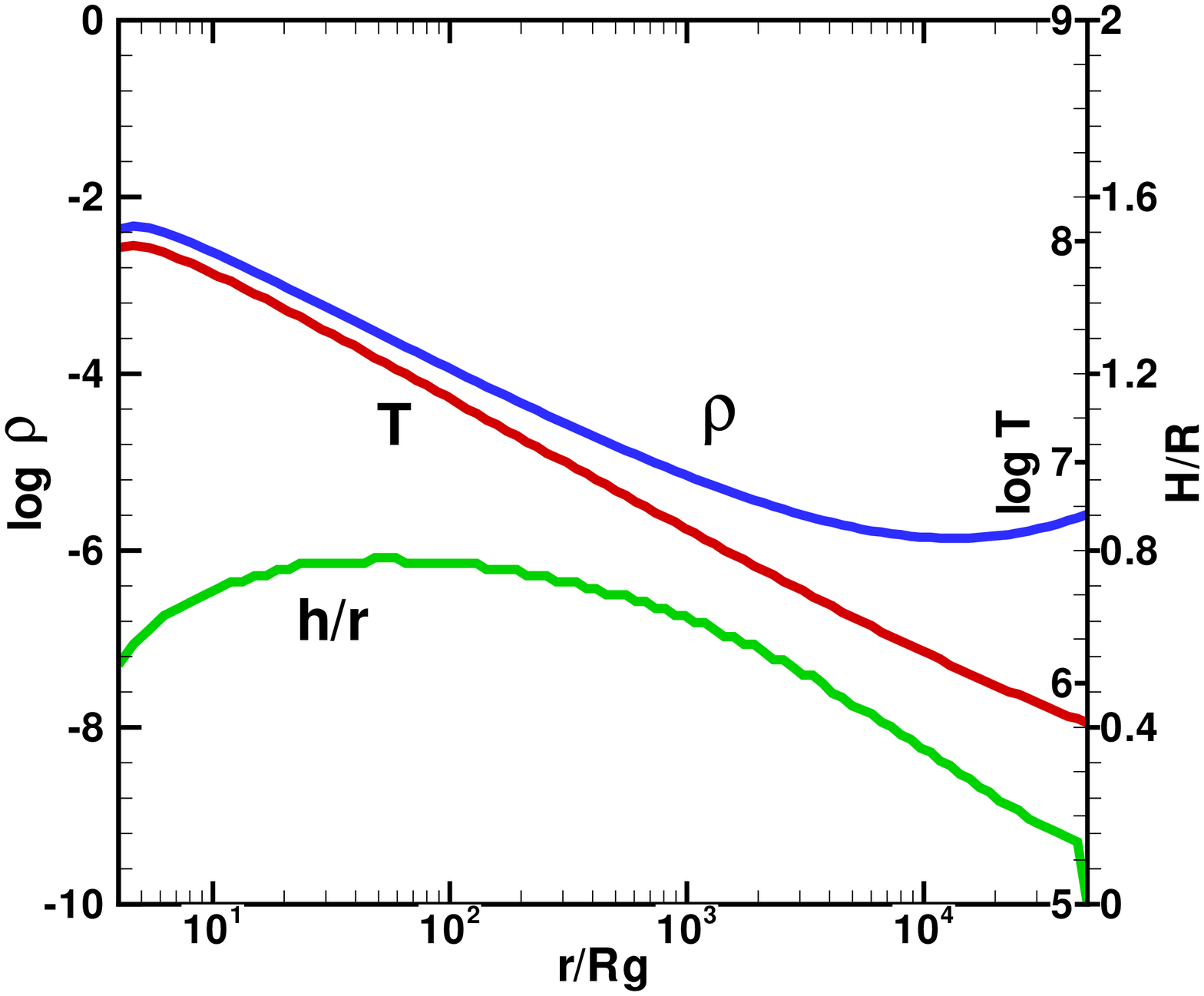}
\end{center}
\caption{Disc thickness $h/r$, temperature $T$ (K), and density $\rho$ (g$\;{\rm cm}^{-3}$) of the supercritical disc with advection 
 for the case of  $\dot m_0 \sim 600$ and $\alpha=0.1$.
}
\label{fig:fig3}
\end{figure}

\begin{table*}
\centering
\caption{Results and comparison with observations}
\begin{tabular}{@{}ccccccccc} \hline \hline
 ${\rm Model}$ & $\alpha$ & $  \dot M_0 $ & 
 $ L $ & $L_{\rm d} $ & 
 $  \dot M_{\rm out} $ & $  \dot M_{1^\circ}$
  & $ \dot M_{\rm edge}$ & $t_{\rm ev}$ \\
  & & $({\rm M_{\odot}}{\rm \;yr}^{-1})$ &$({\rm erg \;s}^{-1})$ 
  &$({\rm erg\; s}^{-1})$ & $({\rm M_{\odot}}\;{\rm yr}^{-1})$ &
   $({\rm M_{\odot}}\;{\rm yr}^{-1})$ 
       & $({\rm M_{\odot}}\;{\rm yr}^{-1})$ &($R_{\rm g}/c)$      \\\hline
  ML-1 &$10^{-3}$            &$1.8\times 10^{-4}$     &$1.7\times 10^{40}$ 
       &$2.2\times 10^{40}$  &$4.5\times 10^{-6}$   & $1.6\times 10^{-7}$  
       &$2.2\times 10^{-5}$ & $3.3\times 10^5$  \\
  ML-2 &0.1                  &$1.8\times 10^{-4}$     & $1.0\times 10^{40}$
       &$1.1\times 10^{40}$  &$1.8\times 10^{-5}$   & $4.7\times 10^{-9}$ 
       &$1.2\times 10^{-7}$  & $5.2 \times 10^5$ \\
  AD-1 &$10^{-3}$            &$1.8\times 10^{-4}$   & $2.5\times 10^{40}$ 
       &$3.2\times 10^{40}$  & $3.8 \times 10^{-5}$ & $9.2 \times 10^{-7}$ 
       & $2.7\times 10^{-5}$ & $ 2.3 \times 10^5$ \\
  AD-2 &0.1                  &$1.8\times 10^{-4}$   & $2.0\times 10^{40}$ 
       & $2.5\times 10^{40}$  & $1.3 \times 10^{-4}$ & $1.2\times 10^{-6}$ 
       &$1.5\times 10^{-5}$  &$1.1 \times 10^6$   \\\hline
  Observation & -- & $10^{-4}$ -- $10^{-3} $ &$10^{39}$ -- $10^{40}$
       & -- &$10^{-5}$ -- $10^{-4}$& $10^{-7}$ -- $10^{-6} $ 
       & -- & --      \\\hline
\end{tabular}
\end{table*}

\begin{figure*}
\begin{center}
 \includegraphics[width=10cm,height=6cm]{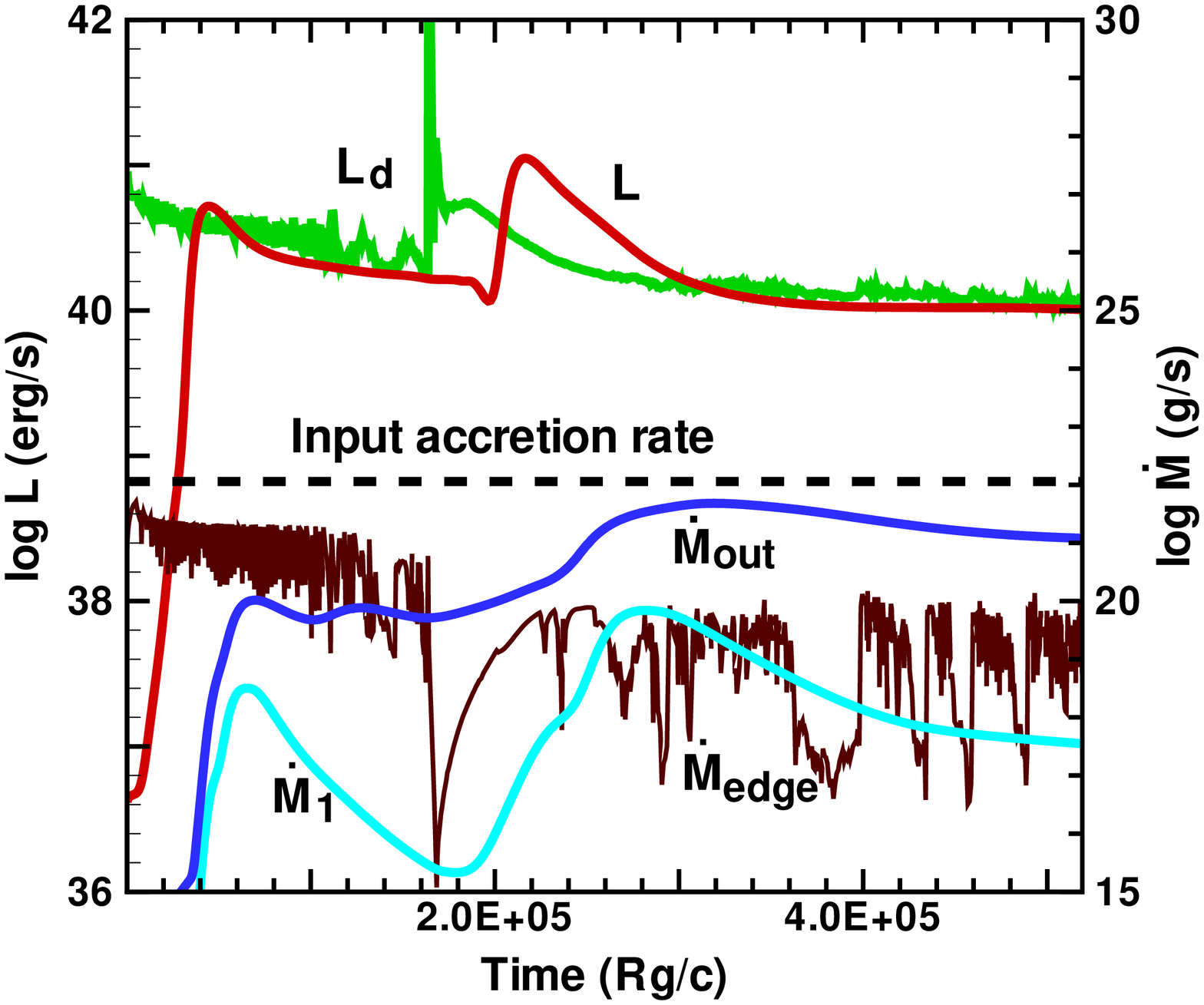}
\end{center}
 \caption{Time evolution of total luminosity $L$, disc luminosity $L_{\rm d}$,
       total mass-outflow rate $\dot M_{\rm out}$ ejected from the entire outer
       boundary surface, mass-outflow rate $\dot 
       M_{1^\circ}$ of the axial outflow with a half opening angle $1.2^\circ$
       along the rotational axis, and mass-inflow rate $\dot M_{\rm edge}$
       into the black hole through the inner boundary for model ML-2, where
       time is shown in units of $R_{\rm g}/c$.
    }
 \label{fig:fig4}
\end{figure*}

\section{Supercritical accretion disc models with mass loss and advection}
 Our previous studies of SS 433 \citep{b28,b30} are  based on 
 the initial discs by S-S model with $\dot m_0 \le 20$.  
  Here, $\dot m_0$ is the input accretion rate normalized to the
 Eddington critical accretion rate $\dot M_{\rm c} (= L_{\rm E}/\eta c^2
 = 4 \pi GM/\eta c \kappa)$, where $L_{\rm E}$ is the Eddington luminosity,
 $G$ the gravitational constant, $M$ the black hole mass, $\kappa$ the 
 Thomoson scattering opacity,  $c$ the velocity of light, and $\eta$
 the conversion efficiency of the energy of the accreting matter into
 radiation.
 Adopting $\eta= 1/12$, we have $\dot M_{\rm c} = 2.0 \times 10^{19}$
 g s$^{-1}$ and $L_{\rm E} = 1.5\times 10^{39}$ erg s$^{-1}$ for 
 $M=10 {\rm M_\odot}$.
 In this paper we consider $\dot m_0\sim 600$, which corresponds
 to $\dot M_0 \sim 1.8 \times 10^{-4} {\rm M_{\odot}} \; {\rm yr}^{-1}$ 
 and is a more plausible value for SS 433 \citep{b8}.
 The S-S disc with such a high accretion rate  is
 geometrically too thick and becomes invalid against assumptions used.
 When the accretion rate is very high, the disc luminosity exceeds the 
 Eddington luminosity and matter flows out of 
 the disc surface, causing the accretion rate onto the central object 
 to decrease. The outflow takes place inside the radius at which the disc
 thickness becomes comparable to the disc radius. This radius is called
 a spherization radius $R_{\rm sp}$ \citep{b35}.
 \citet{b20} proposed supercritical disc models,
 taking account of mass loss and advection through the accretion discs,
 which were applied to SS 433 and the ultraluminous X-ray sources (ULXs)
 by \citet{b33}. 
 In their models with mass loss, it is assumed that a fraction 
 $\epsilon_{\rm W}$ of the radiation energy flux is spent on the production of 
 the outflow within the spherization radius. 
 For $\epsilon _{\rm W} =1$, if advection is neglected,  we obtain analytical
 solutions \citep{b20} for the disc variables at the disc plane, and 
 the ratio $R_{\rm sp}/3R_{\rm g}$ is estimated to be $\sim  {5 \over 3} \dot m_0$, where $R_{\rm g}$ is the 
 Schwarzschild radius.
 The accretion rate in the disc within the spherization radius is given by

  \begin{equation}
    \dot M(r)=\dot M_0 \;\left({r\over R_{\rm sp} }\right)\; {{1+{2\over 3} ({r\over
 3 R_{\rm g}})^
     {-5/2}} \over {1+{2\over 3}({R_{\rm sp}\over 3R_{\rm g}})^{-5/2}}} 
     \;\; {\rm for} \;r \le R_{\rm sp}.
 \end{equation}
 This law has the asymptotic form
 \begin{equation}
  \dot M(r) \approx \dot M_0\; {r\over R_{\rm sp}} \; \; {\rm for} \;
   3R_{\rm g} \ll r \le R_{\rm sp}.
 \end{equation}
The temperature $T$ and the density $\rho$ in the central plane of the disc
 are approximately given 
 in the region of $ 3R_{\rm g} \ll r \leq R_{\rm sp}$  by
  \begin{equation}
  T(r) = 6.4\times 10^7 \left({M\over 10{\rm M_{\odot}}}\right)^{-1/4}\left({\alpha \over 0.001}\right)^
         {-1/4} \left({r\over 3R_{\rm g}}\right)^{-3/8} {\rm K},
  \end{equation}
  \begin{equation}
  \rho(r)=1.4 \times10^{-3}\left({M\over 10{\rm M_{\odot}}}\right)^{-1}\left({\alpha\over 0.001}\right)^{-1} \left({r \over 3R_{\rm g}}\right)^{-1/2} {\rm g \;cm}^{-3},
  \end{equation}
where $\alpha$ is the viscosity parameter.

  Figs 1 and 2 show the accretion rate $\dot m(r)$, the relative disc 
 thickness $h/r$, the central plane temperature $T$, and the density 
 $\rho$ for the 1D model with mass loss and the viscosity parameter 
 $\alpha=0.1$, along with the solution for S-S model without mass loss.
 Outside the spherization radius the solutions are identical.

 The disc at the supercritical accretion rates becomes
 geometrically thick. The emission from the surface of a thick disc is 
 not an efficient cooling mechanism, and the advective transport of the 
 viscously generated heat becomes important in the energy balance 
 equation \citep{b31}. An advection-dominated disc model at moderately 
 super-Eddington accretion rates $\dot m_0 \le 70$ was first proposed by
 \citet{b1}.
  A supercritical disc with even higher accretion rates was approached 
 by \citet{b20}, where mass loss and advection were included, and the 
 discs were geometrically thick ($h/r \sim 0.7$ at maximum) but not so
 thick as it followed from S-S model ($h/r \ge 100$).
 Fig. 3 shows a solution for the advective supercritical disc with mass
 loss for  $\dot m_0 \sim 600$ and $\alpha=0.1$.  
 This solution and the analytic solution for the disc with mass loss,
 described above, are used in the current work as initial configurations
 of the disc in the 2D numerical model.

\section{Numerical Methods}

 A set of relevant equations for the numerical calculation consists of
  six partial differential 
 equations for the density, the momentum, and the thermal and radiation energy.
 These equations include the full viscous stress tensor, the heating and 
 cooling of the gas, and the radiation transport. 
 The pseudo-Newtonian potential \citep{b32} is adopted in the momentum
 equation. 
 The radiation transport is treated in the gray, flux-limited diffusion
 approximation \citep{b18}.
 We  use spherical polar coordinates ($r$,$\zeta$,$\varphi$), where $r$
 is the radial distance, $\zeta$ is the angular distance measured from 
 the equatorial plane of the disc, and $\varphi$ is the azimuthal angle.
  The above set of differential equations is numerically integrated in time
 using a finite-difference method, which is an improved version of that 
 described in \citet{b15}. 
 The method is based on an explicit-implicit finite difference scheme, whose
 details are described by \citet{b29} and \citet{b30}.
 The computational domain is divided into $N_r \times N_\zeta$ grid cells,
 where $N_r$ grid points (=100) in the radial direction are spaced 
 logarithmically as $\Delta r/r = 0.1$ and $N_\zeta$ grid points (=150) 
 in the angular direction are equally spaced, but more refined near the
 equatorial plane, typically with $\Delta \zeta=
 \pi/150$ for $\pi/2 \geq \zeta \geq \pi/6 $ and $\Delta \zeta=
 \pi/300$ for $\pi/6 \geq \zeta \geq 0 $, in order to resolve the structure
 of the accretion disc.
 Although the radial mesh-sizes do not have a fine 
 resolution to examine detailed disc structure, we consider the mesh-sizes to
 be sufficient for examination of the global behavior of the disc, the jets, 
 and the wind mass-outflow rates.

 \subsection{Model Parameters}
 We consider a Schwarzschild black hole with mass $M= 10 {\rm M_{\odot}}$
 and take the inner-boundary radius $R_{\rm in}= 2 R_{\rm g}$ and the outer boundary radius of the spherical computational domain $R_{\rm out}= 4.5\times 10^4 R_{\rm g} (\sim 1.4\times 10^{11}$ cm). 
 For the kinematic viscosity, we adopt the usual $\alpha$ - model.
 In Table 1, we give the parameters of the discs with mass loss (ML) and 
 advection (AD),
 where $\dot M_0$ is the input accretion rate, and $\alpha$ is the 
 viscosity parameter. 
 Starting with 1D solutions described in section 2, we perform time 
 integration of the equations until a quasi-steady solution is obtained.

\subsection
{ Initial Conditions}
 The initial conditions  consist of a dense, optically thick accretion disc
 and a rarefied optically thin atmosphere around the disc. 
 Physical variables $\rho$ and $T$ at the equatorial plane for $r \ge 
 3 R_{\rm g}$ are given by the 1D solutions for the 
 supercritical discs with mass loss or advection.
 We construct the vertical structure of the disc in the approximation of
 the hydrostatic and radiative equilibrium by integrating the
 relevant equations with given boundary values at the equatorial plane.
 As for the initial atmosphere around the disc,
  assuming that the gas is in the optically thin limit and in the radially 
 hydrostatic equilibrium,
 we have
  \begin{equation}
   F_{\rm r}= c E_{\rm r} = -{\lambda c\over \rho\kappa_{\rm e}}
        {\partial E_{\rm r}\over\partial r}, 
  \end{equation}
  \begin{equation}
   {\partial P\over\partial r} = -\rho {GM\over r^2},
  \end{equation}
  where $F_{\rm r}$ is the radial component of the radiative flux ${\bf F}$,
  $E_{\rm r}$ the radiation energy density, $\lambda$ the
  flux-limiter, $\kappa_{\rm e}$ the electron-scattering opacity, and $P$ 
 the total pressure. Furthermore, if it is assumed that the flux-limiter
 $\lambda$ is constant and that the radiation pressure is dominant throughout
 the gas, we have for the initial radiation energy density $E_{\rm r}$ and
 the density $\rho$, from the above equations,
  \begin{equation}
   E_{\rm r} = {1\over\kappa_{\rm e}} {GM\over r^2},
  \end{equation}
  \begin{equation}
   \rho = {2\lambda\over {\kappa_{\rm e} r}}.
  \end{equation}
  Actually, the flux-limiter $\lambda$ is taken to be $\sim 10^{-3}$.
  However, we note that a particular initial distribution of the gas
 around the disc does not influence the results at the sufficiently late 
 simulation times.

\subsection {Boundary Conditions}

   Physical variables at the inner boundary, except for the velocities,
    are calculated by extrapolation of the variables near the boundary. 
  We impose limiting conditions that the radial velocities at the inner 
 boundary are given by a fixed free-fall velocity and the angular 
 velocities are zero.
   On the rotational axis and the equatorial plane, 
 the meridional tangential velocity $w$ 
 is zero and all scalar variables must be symmetric relative to the axis
 and the plane.
 The outer boundary at $r=R_{\rm out}$ is divided into two parts. 
 One is the disc boundary through which matter is entering from 
 the outer disc.
 At this outer-disc boundary we assume a continuous inflow of matter 
 with a constant accretion rate $\dot M_0$.   
 The other part is the outer boundary region above the
 accretion disc. We impose free-floating conditions on this outer boundary 
 and allow for outflow of matter, whereas any inflow is prohibited here. 
 We also assume the outer boundary region above the disc to be 
 in the optically-thin limit, 
 $\vert \mbox{\boldmath$F$} \vert \rightarrow c E_{\rm r}$.

 \begin{figure}
 \begin{center}
 \includegraphics[width=8cm, height=6cm]{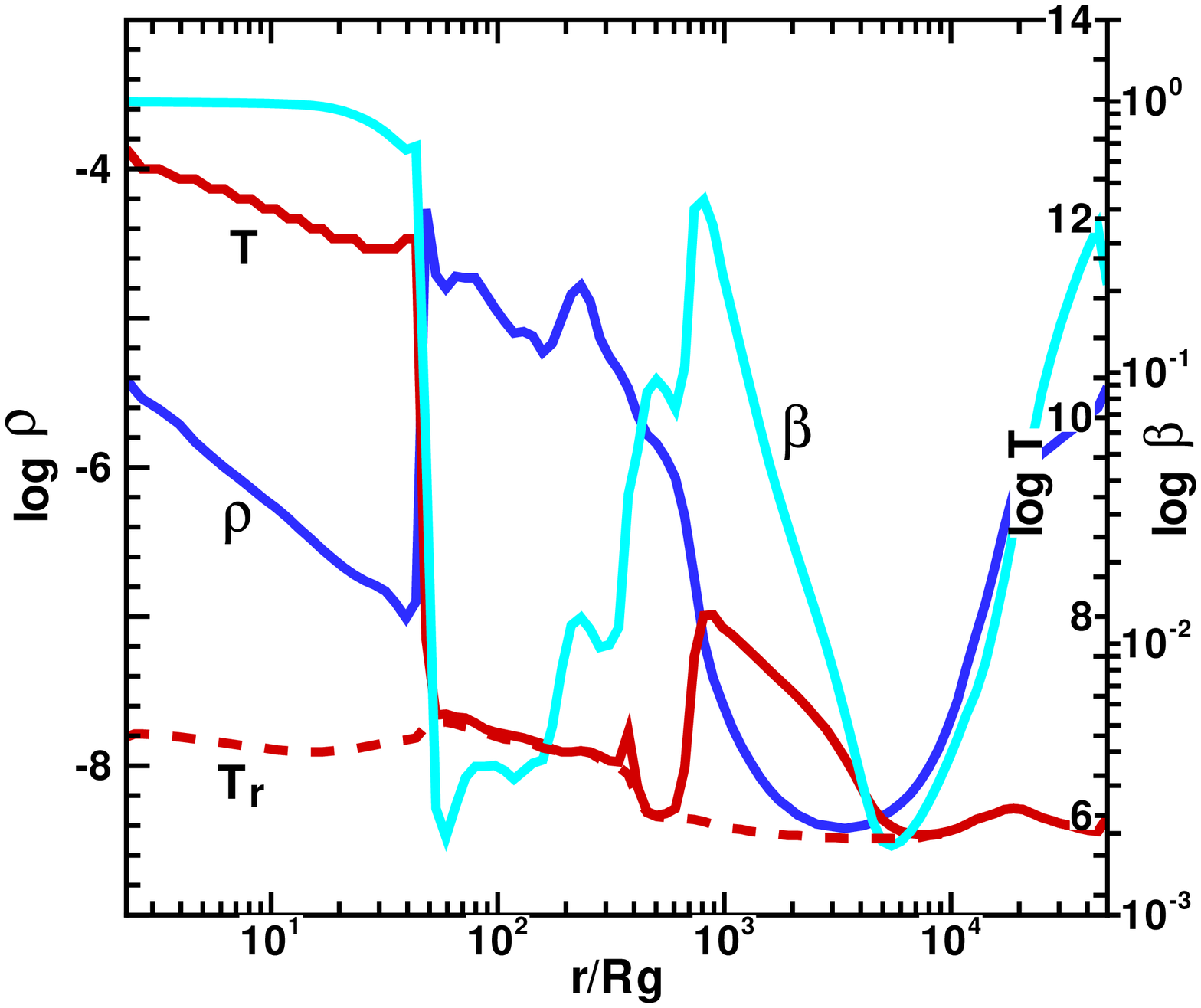}
 \end{center}
 \caption {Radial profiles of density $\rho$ (g ${\rm cm}^{-3}$),  
   temperature $T$ (K), radiation temperature $T_{\rm r}$ (K) (dashed line),
@ and ratio $\beta$ '†of the gas pressure to the total pressure on the 
  equatorial plane at $t=5.2\times 10^5 R_{\rm g}/c$ for model ML-2.
 }
 \label{fig:fig5}
 \end{figure}

\section{Numerical Results}
 
 In order to obtain a reliable configuration of the jet, the simulation time
 should be longer than the jet arrival time
  $t_{\rm jet}$ at the outer boundary. 
 On the other hand, to get a steady state of a viscous disc in a
 region of radius $r$,
 we need  the simulation time longer than the characteristic dynamical 
 time $t_{\rm dyn}$ and the viscous time-scale $t_{\rm vis}$.
 These times are given as follows:
  \begin{equation}
   t_{\rm jet} \sim {R_{\rm out}\over V_{\rm jet}} = 2.2\times 10^5
     \left({V_{\rm jet}\over 0.2c}\right)^{-1} {R_{\rm g}\over c},
 \end{equation}
  \begin{equation}
    t_{\rm dyn} \sim {\Omega}^{-1} = \left({r\over R_{\rm g}}\right)^{3/2}
      {R_{\rm g}\over c}
    = 3\times 10^4 \left({r\over 10^3 R_{\rm g}}\right)^{3/2} 
    {R_{\rm g}\over c}, 
  \end{equation}
and
  \begin{equation}
    t_{\rm vis} \sim {t_{\rm dyn}\over \alpha} {\left({h\over r}\right)}^{-2} 
    = 3\times 10^5 {\left({\alpha\over 0.1}\right)}^{-1} 
     {\left({h\over r}\right)}^{-2}
     {\left({r \over 10^3R_{\rm g}}\right)}^{3/2} {R_{\rm g}\over c},
    \end{equation}
where $\Omega$ and $V_{\rm jet}$ are the Keplerian angular velocity and the
typical jet velocity.
 For the supercritical disc of SS 433, we have $h/r \sim 1$ and
 $V_{jet} \sim 0.27c$. Large-scale mass loss  sets in near the spherization
 radius of the disc at $r \sim 10^3 R_{\rm g}$. For this radius and 
 $\alpha =0.1$, the maximum time of $t_{\rm jet}$, $t_{\rm dyn}$,
 and $t_{\rm vis}$ is $\sim 3\times 10^5 R_{\rm g}/c$. 
 We set this time as an approximate computational goal, exceeding it in
 the cases of models ML-2 and AD-2 in order to examine the disc instability.
 Whereas the quasi-steady values of the luminosities are almost attained
 at the final phases,  the simulation times are still not  
 sufficient for the disc and the outflow to settle into a completely steady
 state, because the input accretion rate is not equal to the total
 mass-outflow rate plus the mass-flux rate of the falling gas into the black
 hole.
 In spite of the limited computational time, 
 we are able to derive the general properties of the disc and the outflow. 
 
 The luminosity curve is a good measure to check if a steady state of the disk
 and the outflow is attained. The total luminosity $L$  and the disc 
 luminosity $L_{\rm d}$  are calculated as $\int {\bf F} d {\bf S}$,
 where the surface integral is taken over the outer boundary surface and
 the disc surface, respectively. 
 The disc surface is placed at the height where the density
 drops to a tenth part of the value at the equatorial plane. This lead to
 some uncertainty in $L_{\rm d}$ because the vertical structure of a 
 geometrically thick disc is treated rather approximately.

 To compare with the observational data for SS 433, we calculte the 
 total mass-outflow rate $\dot M_{\rm out}$ and the ``axial outflow'' rate
 $\dot M_{1^\circ}$. The value $\dot M_{\rm out}$ is calculated for the 
 entire outer-boundary surface and corresponds to the observed mass-outflow
 in the wind of SS 433. Observations indicate that the opening angle of the
 X-ray and optical jets is  $ \sim 1.2^{\circ}$ \citep{b23,b8}.
 Throughout the paper, we call the relativistic outflow through the outer
 base of the cone directed along the rotational axis with  the half
 opening angle of $ 1.2^{\circ}$  ``the axial outflow''. 
 The mass-flux rate $\dot M_{1^\circ}$  of the axial outflow should 
 correspond to the mass-outflow rate in the jets of SS 433, 
 provided the size of the present computational domain is comparable to
 the distance of the observed jets from the central source.
 In order to check the total mass-flux rate, we also calculate the mass-flux
 rate $\dot M_{\rm edge}$ of the gas falling through the inner boundary
 into the black hole.

 In Table 1 we give $L$, $L_{\rm d}$, $\dot M_{\rm out}$, $\dot M_{1^\circ}$,   and $\dot M_{\rm edge}$ at the final simulation time $t_{\rm ev}$ 
 for each of  the models.
 Values $L$, $L_{\rm d}$, and $\dot M_{\rm edge}$ show QPOs-like
 features in model AD-2; Table 1 shows their averaged values around
  the final phase. In the columns for $\dot M_{\rm out}$ and $\dot M_{1^\circ}$,  one can see the observational mass-outflow rate in the wind and the jets 
 of SS 433 \citep{b5,b36,b16,b23,b4,b8}.

\begin{figure}
 \begin{center}
 \includegraphics[width=8cm, height=6cm]{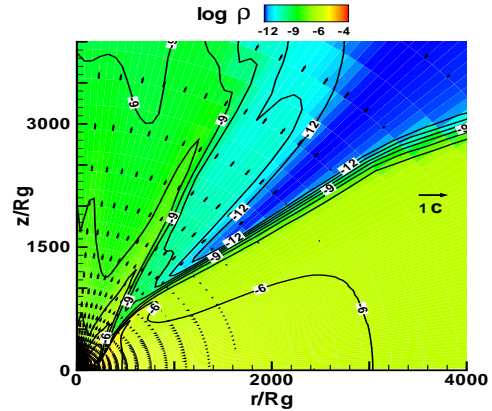}
 \end{center}
 \caption {Velocity vectors and density contours near the spherization 
 radius on the meridional plane at $t = 3.2 \times 10^4 R_{\rm g}/c$
  for model ML-2. 
   The reference vector of light speed is shown by a long arrow.
  The axial outflow propagate with relativistic velocities $\sim 0.2 c$
   along the Z-axis.
 }
 \label{fig:fig6}
 \end{figure}

\begin{figure}
 \begin{center}
\includegraphics[width=8cm, height=6cm]{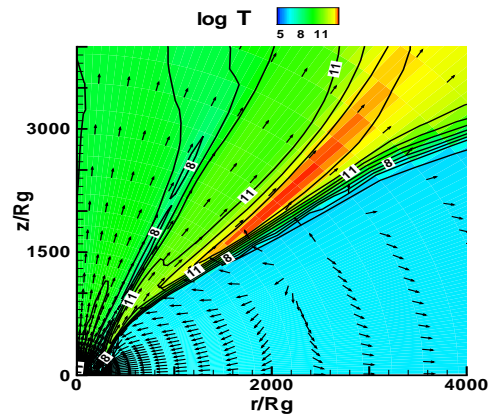}
 \end{center}
 \caption { Same as Fig.6 but for temperature contours. The velocity
             vectors are indicated by unit vectors. 
 }
 \label{fig:fig7}
 \end{figure}

\subsection{Cases of ML-1 and ML-2 with mass loss}

 \begin{figure}
 \begin{center}
 \includegraphics[width=8cm, height=6cm]{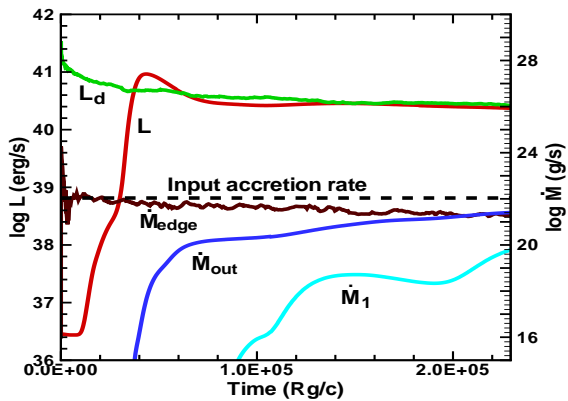}
 \end{center}
 \caption {Time evolutions of $L$, $L_{\rm d}$, $\dot M_{\rm out}$, 
       $\dot M_{1^\circ}$, and  $\dot M_{\rm edge}$  for model AD-1.
 }
 \label{fig:fig8}
 \end{figure}

 In the model ML-1, the initial disc is radiation-pressure dominant and
  optically thick. 
 The perturbed density and temperature waves are generated initially near the
 inner edge, and then  heating and cooling waves move up and down between
 the inner edge and the radius of $\sim 8R_{\rm g}$, 
 raising the disc temperature
 to $\sim 10^{12}$ K. As a result, the gas-pressure dominat, optically thin
 region is formed in the innermost region. 
 The process is repeated for a while, then stops after
 $t \sim 4 \times 10^4 R_{\rm g}/c$. 
 Finally, we get stable, optically thick, and radiation-pressure dominant
 disc. 
 The resultant disc temperatures at the equatorial plane are not different 
 significantly from the initial temperatures of the disc, while the densities
 are enhanced by a factor of 10 -- 30 in the region of $r \le 10^2 R_{\rm g}$.
 The final luminosities $L$ and $L_{\rm d}\sim 2 \times 10^{40}$
  erg s$^{-1}$ are one order of magnitude larger than the Eddington
 luminosity. The total mass-outflow rate $\dot M_{\rm out} \sim 4.5\times 
 10^{-6}{\rm M_{\odot}}\;{\rm yr}^{-1}$ is less by an order than the 
 mass-outflow rate of the wind observed in SS 433, and the rate of the axial
 outflow $\dot M_{1^\circ} \sim 1.6 \times 10^{-7}{\rm M_{\odot}}\;
 {\rm yr}^{-1}$ is marginally in the range of the observed mass-outflow 
 rate of the jets.

 Fig. 4 shows the time evolutions of $L$, $L_{\rm d}$, $\dot M_{\rm out}$, 
 $\dot M_{1^\circ}$, and $\dot M_{\rm edge}$  for model ML-2 with the viscosity
 parameter $\alpha$=0.1.
 The total luminosity $L$  becomes comparable to the disc luminosity 
 $L_{\rm d}$ at the time $t \sim R_{\rm out}/c = 4.5 \times 
10^4 R_{\rm g}/c$, which is the photon transit time from the inner edge 
to the outer boundary.  
 The total luminosity $L \sim 10^{40}$ erg s$^{-1}$ obtained finally is 
 greater by a factor of 6 than the Eddington luminosity. 
 
 The initial disc of ML-2 is radiation-pressure dominant and optically
 thick throughout the whole disc region.
 In the same way as in the case of ML-1, the perturbations of density and 
 temperature, generated initially near the inner edge, propagate 
 outward and inward as heating and cooling waves, and the high temperature,
 gas-pressure dominant, optically thin region is formed in the inner 
 disc. 
 The instability in the optically thin region is
 never damped, as indicated by  $\dot M_{\rm edge}$ in Fig. 4,
 and appears as an oscillating hot blob with a variable size of  
 10 -- 100 $R_{\rm g}$ at the equatorial plane.
  After a large-scale hot blob at $t \sim 1.6
 \times 10^5 R_{\rm g}/c$, the instability persists but the absolute variation
 amplitude of $\dot M_{\rm edge}$ becomes smaller.
 In spite of the considerable variability of $\dot M_{\rm edge}$, 
 the modulations of $L$ and $L_{\rm d}$ are weak, except for the phase of
 the large-scale blob, and become negligible at the later phase.
 At the final phase, the first outward heating wave reaches the 
 spherization radius and merges into the outer Shakura-Sunyaev disc, and 
 the whole disc tends to settle gradually into a steady state.  
 Eventually, the disc evolves into two zones: the gas-pressure dominant
 and optically thin inner disc
 and the radiation-pressure dominant and optically thick outer disc.

\begin{figure*}
 \begin{center}
 \includegraphics[width=14cm, height=6cm]{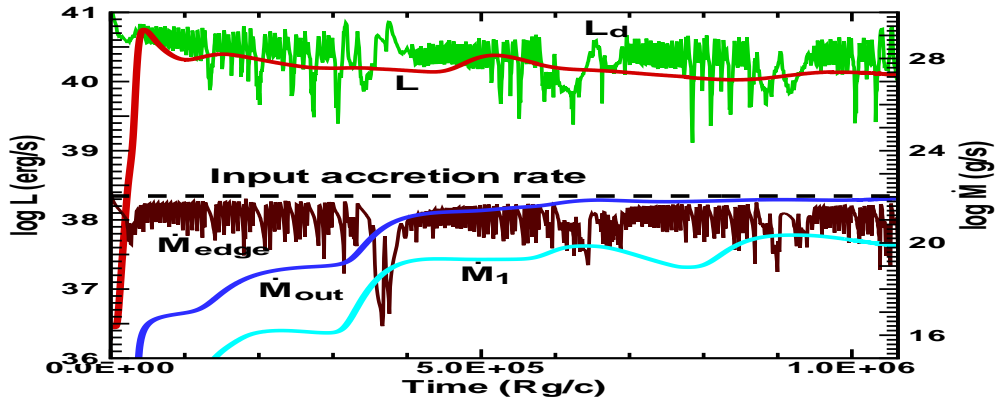}
 \end{center}
 \caption {
   Time evolutions of $L$, $L_{\rm d}$, $\dot M_{\rm out}$, 
       $\dot M_{1^\circ}$, and  $\dot M_{\rm edge}$  for model AD-2.  
       The disc luminosity and the mass accretion rate at the inner edge
       show two types of variability: one with the small amplitude and
       short periods and another with large amplitude and long periods.
   Only the longer-term variations influence the total luminosity in the form
   of small modulations measured at the distant outer boundary.
 }
 \label{fig:fig9}
 \end{figure*}

 Fig. 5 shows the radial profiles of density $\rho$ (g ${\rm cm}^{-3}$),
 gas temperature $T$ (K), radiation temperature $T_{\rm r}$ (K), and 
 ratio $\beta$ of the gas pressure to the total 
 pressure on the equatorial plane at 
 $t=5.2\times 10^5 R_{\rm g}/c$ for model ML-2.
 The radiation temperature $T_{\rm r}$ is defined as $ T_r = 
 (E_{\rm r}/a)^{1/4}$, where $a$ is the radiation density constant.
  When the gas is fully optically thick, the gas temperature $T$ 
 should be equal to the radiation temperature $T_{\rm r}$. 
 When the gas becomes optically thin, $T_{\rm r}$ is much lower
 than $T$ as it is found in Fig. 5.
 Initially, the radiation temperature $T_{\rm r}$ and the gas temperature
 $T$ are $\sim 10^7$ K near the inner edge of the optically thick disc. 
 However, after some time, the gas temperature near the inner edge 
 goes up to $ 10^{10}$ -- $10^{12}$ K and is variable by a factor of 10. 
 Similarly, the radiation temperature is also modulated between 
 $ 10^6$ -- $10^7$  K in the region of $r \le 100 R_{\rm g}$. 
  The gas-pressure dominant, hot, optically thin inner disc
 is separated from the radiation-pressure dominant outer disc by 
 the transition region at $ r \le 100 R_{\rm g}$. 
 The gas variables  $T$ and $\rho$ vary sharply across the transition region,
 but the radiation temperature $T_{\rm r}$ is smooth everywhere.

  Figs 6 and 7 show the contours of density $\rho$ (g cm$^{-3}$ ) 
  and temperature $T$ (K) with the velocity vectors of the gas near the 
 spherization radius at $t= 3.2\times 10^4 
  R_{\rm g}/c$ for model ML-2.
  Here we see  a rarefied, hot, and optically thin high-velocity outflow 
  region and a dense, cold, and geometrically thick 
   disc. The outflow propagates with relativistic velocities of 
 0.08 -- 0.2$c$ along the Z-axis.  The disc is geometrically thick with
 $h/r \sim$ 0.7 -- 1 in the region of $ 10^3  R_{\rm g} \le r \le 
 4\times 10^3  R_{\rm g}$ and the large-scale outflow sets in the region.
 In the high-velocity region along the rotational axis, the temperature is as
 high as $\sim 10^8$ K and the density is as low as $\sim 10^{-11}$ -- 
 1$0^{-9}$ g cm$^{-3}$.

\subsection{Cases of AD-1 and AD-2 with advection}
 Fig. 8 shows the time evolutions of $L$, $L_{\rm d}$, $\dot M_{\rm out}$,
 $\dot M_{1^\circ}$, and $\dot M_{\rm edge}$ for model AD-1. 
 The evolutionary features of the disc is very similar to the case of ML-1.
  Finally, we get stable, optically thick, and
  radiation-pressure dominant disc with the maximum disc temperature
  $\sim 10^8$ K in the innermost region. 
 The total luminosities $L \sim 2.5 \times 10^{40}$ and $2.0 \times 10^{40}$
 erg s$^{-1}$ in models AD-1 and AD-2 are in the same range as those in 
 models ML-1 and ML-2. 
 The total mass-outflow rates are $\sim 3.8 \times 10^{-5}$
 and $1.3 \times 10^{-4} {\rm M_{\odot}}$ yr$^{-1}$ and the rates of the 
 relativistic axial outflow are
 $9.2 \times 10^{-7}$ and $1.2 \times 10^{-6} {\rm M_{\odot}}$ yr$^{-1}$
 in models AD-1 and AD-2 respectively. These values of $\dot M_{\rm out}$
 and $\dot M_{1^\circ}$ are consistent with the corresponding observed rates
 of the wind and the jets.
 
 Fig. 9 shows the time evolutions of $L$, $L_{\rm d}$, $\dot M_{\rm out}$,
 $\dot M_{1^\circ}$, and $\dot M_{\rm edge}$ for model AD-2.
 In this case, we find the remarkable variabilities of $\dot M_{\rm edge}$,
 $L_{\rm d}$, and $L$ and, to examine their properties, we took a long 
 simulation time $t \sim 10^{6} R_{\rm g}/c$ ($\sim$ 100 s).
  The high-velocity jets propagate vertically to the disc plane and expand 
  gradually from the rotational axis with increasing time.
  After the time $R_{\rm out}/0.2c$, the jets arrive at the outer boundary 
  in the polar direction, and the mass outflow begins and gradually approaches
 the steady state. 
 In model AD-2, most of the accreting matter is flown out as  wind and
 only 10 percent of the input matter is swallowed into the black hole. 
 
 The modulations of $\dot M_{\rm edge}$ in AD-2 show two types of variability:
 (1) the small amplitude variations with a short time-scale,
 (2) the large amplitude variations with a long time-scale.
 The disc luminosity is also modulated by a factor of a few to ten,
 synchronously with $\dot M_{\rm edge}$.
 The variations with the small amplitude and short 
 periods strongly influence the disc luminosity  but not the
 total luminosity measured at the distant outer boundary.
 If the atmosphere between the disc surface and the outer boundary was 
 fully optically thin, the total luminosity would suffer from 
 the same  modulations as the disc luminosity. However, in the supercritical
 accretion flow with high input density, only the large modulations
 with the long periods contribute to the total luminosity, due to the 
 atmospheric absorption around the disc. 
 These modulations behave as QPOs-like variabilities. 
 In Fig. 10, we plot the power density spectra of $\dot M_{\rm edge}$, 
 $L_{\rm d}$, and $L$ and recognize  the QPOs-like features of these 
 spectra with characteristic signals of $\nu \sim 4 \times 10^{-2}$ --
 $10^{-1}$ and 0.5 -- 2 
 Hz for $\dot M_{\rm edge}$ and $L_{\rm d}$.
 Only longer-period variations are obtained for $L$; therefore, we expect
 modulations of the observed luminosity with the quasi-periods between 
 $\sim$ 10 and 25 s.

 \begin{figure}
 \begin{center}
 \includegraphics[width=8cm, height=6cm]{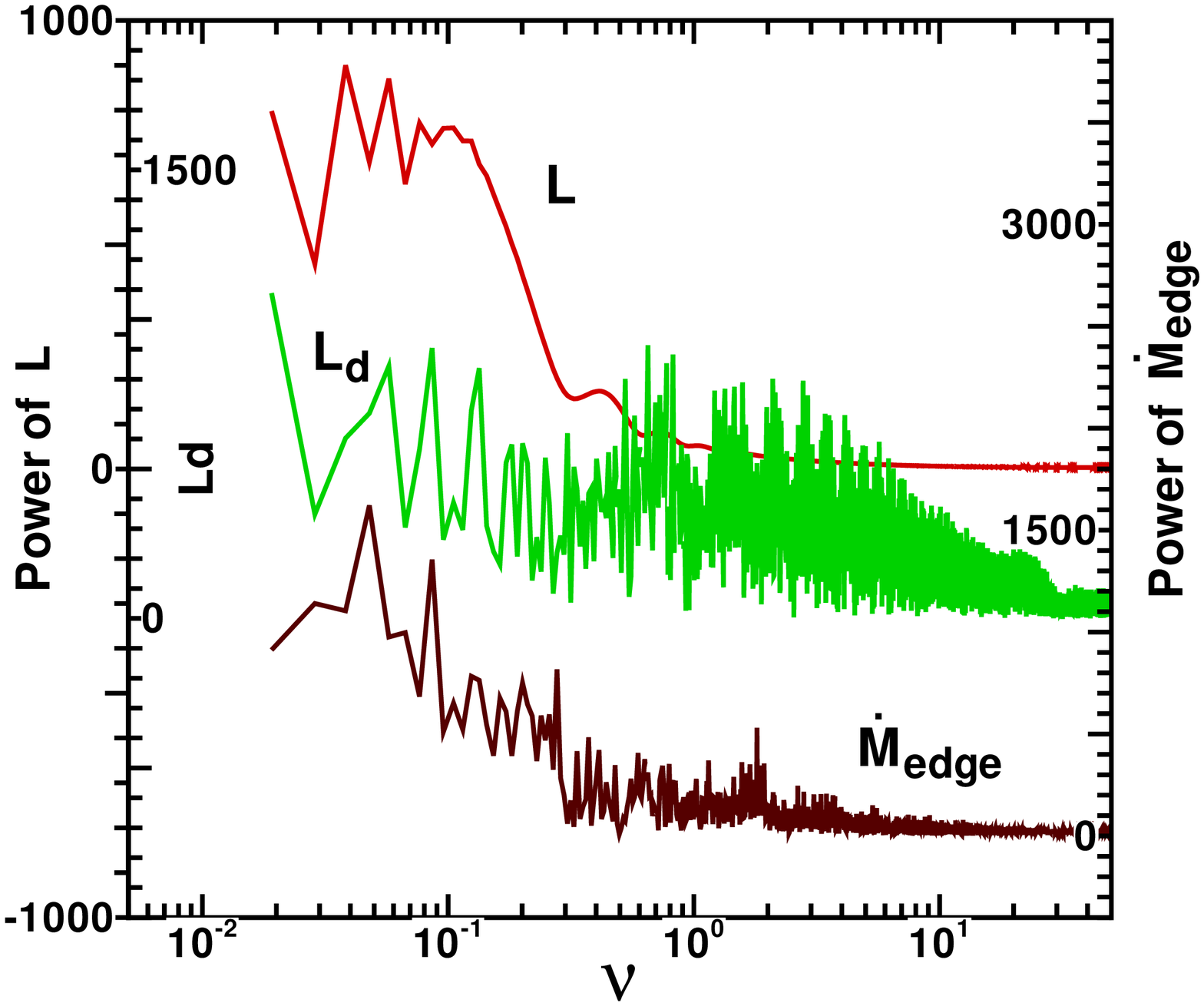}
 \end{center}
 \caption {Power spectra of time variations of the total luminosity 
      $L$, the disc luminosity $L_{\rm d}$, and the mass-inflow rate 
    $\dot M_{\rm edge}$ at the inner edge of the computational domain
 for model AD-2.
 }
 \label{fig:fig10}
 \end{figure}

 \begin{figure}
 \begin{center}
 \includegraphics[width=8cm,height=6cm]{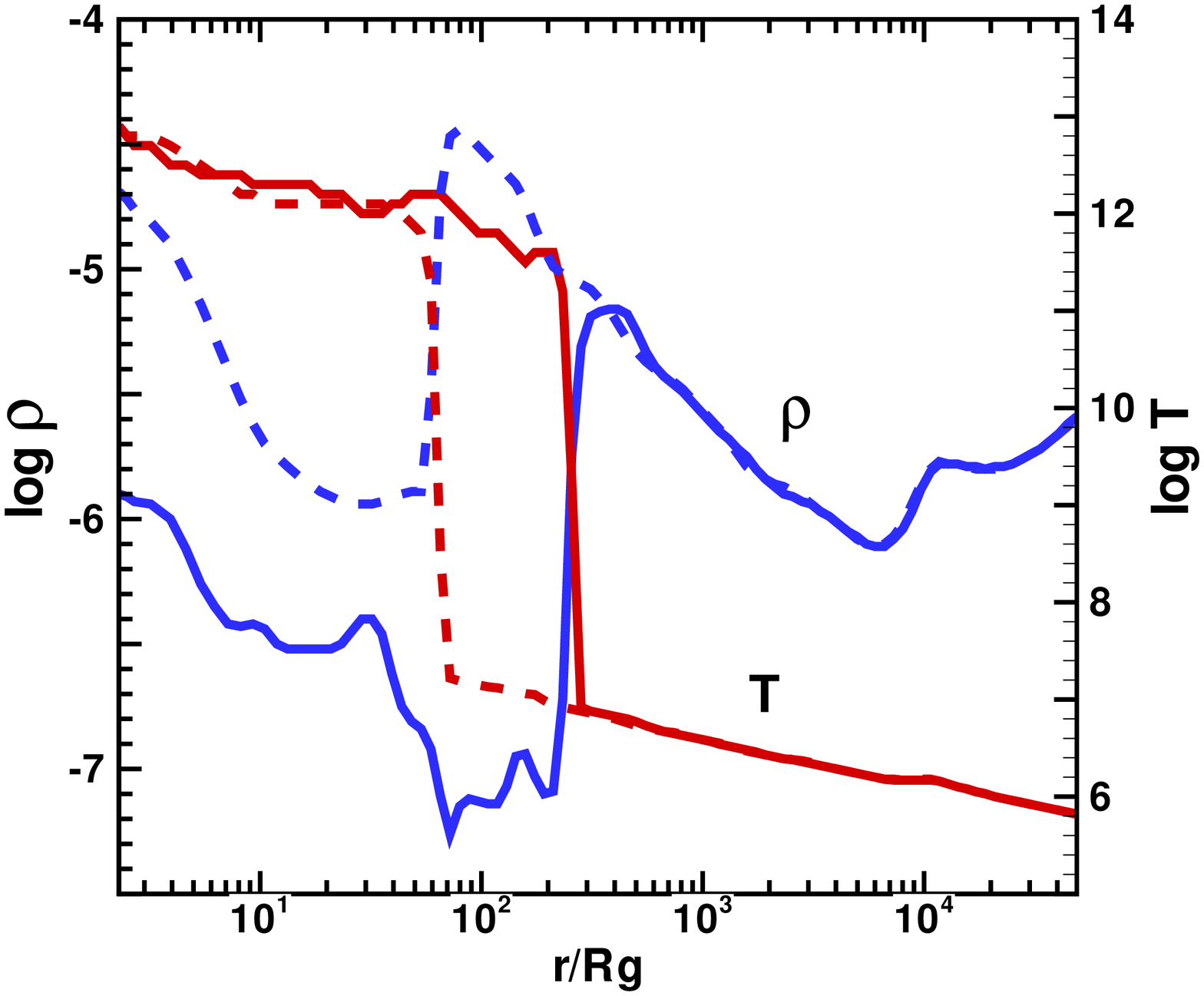}
 \end{center}
 \caption{Radial profiles of density $\rho$ (g $cm^{-3}$) and temperature 
   $T$ (K) at the equatorial plane for AD-2, where dashed and solid lines 
   show the profiles at the phases of $t=1.03 \times 10^6$ and $1.06 
 \times 10^6 R_{\rm g}/c$.  The gas-pressure dominant, optically 
 thin disc in the inner region is separated from the advection-dominated, 
 radiation-pressure dominant, optically thick disc by the critical radius
 $r_{\rm c}$, which moves randomly up and down, usually reaches
 $\sim 60 R_{\rm g}$, and sometimes advances to $\sim$ 100 -- 200 $R_{\rm g}$. 
}        
 \label{fig11}
 \end{figure}

 Fig. 11 shows the radial profiles of density $\rho$ (g ${\rm cm}^{-3}$)
 and temperature $T$ (K) at the equatorial plane 
 for $t= 1.03\times 10^6$ (dashed lines)
 and $1.06\times 10^6 R_{\rm g}/c$ (solid lines) for model AD-2.
 Similarly to the case of model ML-2, the disc obtained in model AD-2
 consists of three regions: (1) the gas-pressure dominant, optically 
 thin disc in the inner region, (2) the advection-dominated, radiation-pressure
 dominant, optically thick disc in the intermediate region, 
 (3) the outer radiation-pressure dominant, optically thick Shakura-Sunyaev 
 disc at $r \ge 10^4 R_{\rm g}$.
 The regions (1) and (2) are sharply separated by the critical radius 
 $r_{\rm c}$. The critical radius moves randomly up and down, and usually
 reaches $\sim 60 R_{\rm g}$ with a heating wave, advancing sometimes 
 to $\sim$ 100 -- 200 $R_{\rm g}$ and rarely to the maximum radius $ \sim
  500R_{\rm g}$, then recedes to the inner edge with a cooling wave. 
 These instabilities in the inner disc lead to recurrent hot blobs 
 with high temperatures and low densities, like bubbles in the
 boilling water.
 The hot blobs typically grow to the size of $\sim 60 R_{\rm g}$ in the disc 
 plane, go up with increasing size, and finally decay  at $z \sim 2000 
 R_{\rm g}$.
 The time evolution of the hot blobs  is shown in Fig. 12.

 \begin{figure*}
 \begin{center}
 \includegraphics[width=14cm,height=8cm]{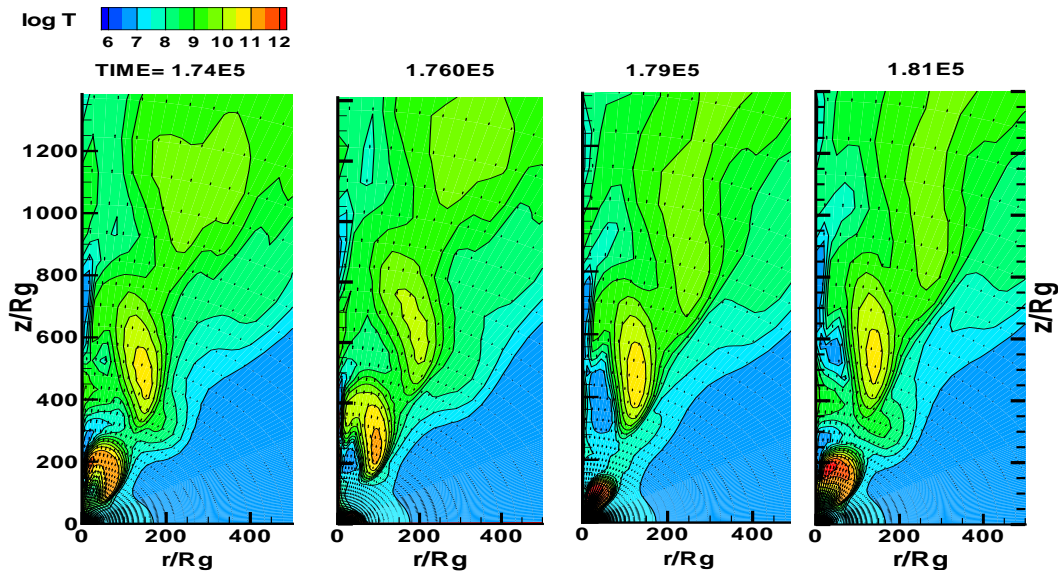}
 \end{center}
 \caption{Time evolution of hot blobs in model AD-2, where 
     the temperature contours of the blobs are shown at $t = 1.74\times 
     10^5$, $1.76 \times 10^5$, $1.79 \times 10^5$, and
     $1.81 \times 10^5 R_{\rm g}/c$.
     The hot blobs typically grow to the size of $\sim 60 R_{\rm g}$ 
     in the disc plane, go up with increasing size, and finally decay
     at $z \sim 2000 R_{\rm g}$.
  }
 \label{fig12}
 \end{figure*}

 Figs 13 and 14 show the contours of gas temperature $T$ 
  (K) and radiation temperature $T_{\rm r}$ (K) with velocity
  vectors in the whole computational domain
  at $t= 1.1\times 10^6 R_{\rm g}/c$ for model AD-2. 
 The gas temperature $T$ in the high-velocity jets region is
 as high as $\sim 10^8$ K at $r \geq 10^{10}$ cm 
 and $T \gg T_{\rm r}$ because the jet region is mildly optically
  thin, but $T \sim T_{\rm r}$ in the outer
 optically thick disc.
 From Fig. 14, we find that the contours of $T_{\rm r}$, as well as
 the contours of radiation energy density $ E_{\rm r}$, show an 
 anisotropic distribution of radial component $F_{\rm r}$ of the 
 radiative flux ${\bf F}$,  where $ {\bf F} \propto - {\rm grad}\;
 E_{\rm r}$.  
 Actually,  the radial components $F_{\rm r}$ in the direction of 
 $\zeta \ge 70^{\circ}$ exceed those in the direction of $\zeta \sim 
30^{\circ}$ by a factor of 5 -- 7.

 Fig. 15 shows the contours of density $\rho\; ({\rm g\; cm}^{-3})$ with 
 magnified velocity vectors over the whole computational domain, where
 the velocity vector of 0.2$c$ is denoted in the legend.
 From detailed analyses of the density contours and the velocity vectors,
 we  recognize three characteristic outflows originating in 
 the different regions of the disc shown in Fig.11:
 (1) the most relativistic axial outflow with velocities $\sim$ 0.1 -- 
  0.3$c$ ejected perpendicularly to the innermost hot, optically thin disc,
 (2) the high-velocity (0.1 -- 0.05$c$) outflow
  within a half opening angle $\sim 30^{\circ}$ ejected from the 
 advection-dominated, optically thick disk in the intermediate region,
 (3) the slow outflow with velocities $\sim 0.01c$ flowing  
 from the disc region near the spherization radius.
 The slow outflow (3) from the outer disc, interacting with the high-velocity
 outflow (2), is  blown obliquely beneath the high-velocity flow
 and is accelerated up to the velocities of 0.002 -- 0.05$c$ for 
 $ 10^{\circ} \le \zeta \le 60^{\circ}$ at the outer boundary.
 The present result for the supercritical disc with $\dot m_0 \sim 600$ 
 shows a broader opening angle of the entire outflow from the disc,
 comparing with $\sim 30^\circ$ in the previous study with $\dot m_0 \sim 20$
 \citep{b30}.
 This is due to the reason that the supercritical disc with a very high
 accretion rate has  a very large spherization radius, where a massive outflow
(3) is going out distortedly from the disc.

\section{Instability of advection-dominated disc}
 The initial discs in all cases considered here are 
 radiation-pressure dominant and optically thick throughout the whole 
 disc region.
 Therefore, their stability may be supposedly interpreted in terms of
 the slim disc model at highly super-Eddington luminosity; that is, 
 these discs should be stable against the local and global perturbations.
 Actually, the time evolution of the disc in models ML-1 and AD-1 with small 
 $\alpha=0.001$ eventually shows stable features.
 On the other hand, just at the beginning of the simulations in ML-2 and AD-2, 
 the unstable behaviors of the accretion rate at the inner edge and 
 the disc luminosity are revealed.
 This can be attributed to the fact that the gas-pressure dominant, optically 
 thin, high temperature state is triggered in the innermost region of
 the disc due to initial perturbations of the disc variables.
 The optically thin disc region, as that developing repeatedly in model AD-2,
 is not attained finally in models ML-1 and AD-1, because the
 disc densities in these models are much higher than those in ML-2 and 
 AD-2 due to $\rho \propto \alpha^{-1}$ in the initial disc. 

 The thermal instability of advection-dominated 
 one-temperature discs was examined by \citet{b13}. It was shown that,
 in the case of $\alpha$ - viscosity, the optically thin advective disc 
 is unstable against local perturbations if the viscosity parameter $\alpha$
 is small, 
  but that two-dimensional analysis is necessary to investigate stability
 if $\alpha$ is large. 
 The stability of an optically thin, advection-dominated 
 accretion disc with large $\alpha = 0.3$ was examined by 1D time-dependent
  numerical simulations \citep{b21}, and it was found that 
 any disturbance added onto the accretion flow at large radii does not
 decay rapidly and continue to be present as  fluctuations in the X-ray
 emission of an accretion disc, though the global disc structure is
 not modified. 
 From the present simulations, we find that the inner, optically thin, 
 gas-pressutre dominant disc in model AD-2 becomes unstable 
 eventually;
 the disc luminosity fluctuates, though the total luminosity at the outer
  boundary does not vary significantly.
 Thus, our two-dimensional calculations of the advection-dominated discs
 confirm the above theoretical and 1D numerical results. 
 On the other hand, for model ML-2 with large $\alpha$,
 we find that $\dot M_{\rm edge}$, $T$, and $T_{\rm r}$ are modulated by a 
 factor of 3 -- 10 in the innermost region of the optically thin disc, 
 that is, the disc is locally unstable.
 However, no significant modulation of the disc luminosity is found 
 especially at the later phases. 
 This is attributed to the fact that, in model ML-2, the absolute value 
  of the local accretion rate in the inner region and its variation amplitude
 are small compared with the input accretion rate. 
 Therefore, the modulations do not influence  the luminosities of $L$ and
 $L_{\rm d}$, because the resultant gravitational energy release due to the
 accreting gas could not contribute largely to the total radiation.

\begin{figure}
 \begin{center}
 \includegraphics[width=8.8cm, height=6.6cm]{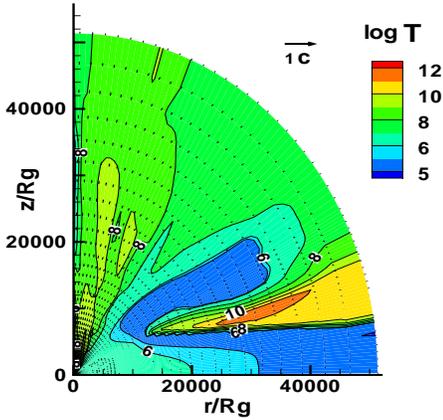}
 \end{center}
 \caption {Velocity vectors and contours of gas temperature $T$ (K) in
        logarithmic
        scale on the meridional plane at $t = 1.1 \times 10^6 R_{\rm g}/c$   
        for model AD-2. 
  }
 \label{fig:fig13}
 \end{figure}

\begin{figure}
 \begin{center}
 \includegraphics[width=8.8cm, height=6.6cm]{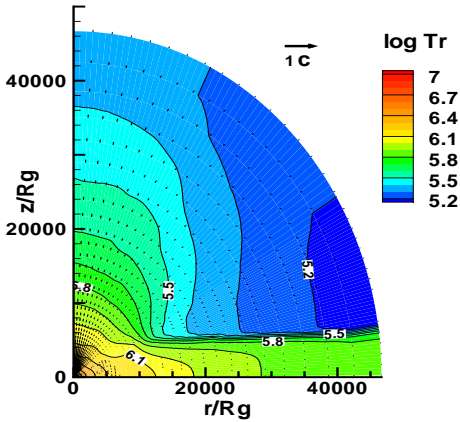}
 \end{center}
 \caption {Same as Fig.13 but for radiation temperature $T_{\rm r}$ (K).
  If the gas is fully optically thick, $T_{\rm r}$ is equal to the 
  gas temperature $T$ (K).
 }
 \label{fig:fig14}
 \end{figure}

\begin{figure}
 \begin{center}
 \includegraphics[width=8.8cm, height=6.6cm]{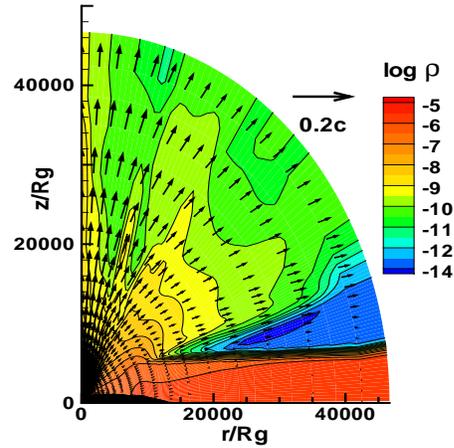}
 \end{center}
 \caption {Same as Fig.13 but for density contours $\rho \;({\rm g \; cm}^{-3})$    with magnified velocity vectors.
 }
 \label{fig:fig15}
 \end{figure}

\section{Comparison with SS 433}
  SS 433 is a typical stellar-mass black-hole candidate with a highly 
 super-Eddington luminosity.
  The present results can be compared with the observations of SS 433,
  because the model input accretion rate $\dot M_0$ ( $\sim 1.8 \times 
  10^{-4} {\rm M_{\odot}} \;{\rm yr}^{-1})$ is in the observed range 
 of SS 433 mass transfer rate \citep{b8}.

 The unique jet velocity $V_{\rm jet}$ (=0.26 $c$) of SS 433 is explained 
 in terms of the relativistic velocity of plasma accelerated by the 
 radiation-pressure force in the inner disc.
 The X-ray spectral lines from the gas moving with relativistic velocities
  are emitted in the very hot, optically thin region along the rotational
  axis. 
 The temperature $T \sim 10^9$ -- $10^7$ K and the density $\rho \sim
 10^{-8}$ -- $7\times 10^{-10}$ g cm$^{-3}$ obtained in the present 
 simulations at distances $6 \times 10^9$ cm $ \le r \le 1.4 \times 10^{11}$
 cm in the conical axial outflow are generally consistent with the values
 $T \sim 1.1 \times 10^8$ -- $6 \times 10^6$ K and $\rho \sim 3 \times
 10^{-9}$ -- $7 \times 10^{-11}$ g cm$^{-3}$, obtained by fitting the model
 of X-ray jet emission lines produced at distances up to $\sim 2
  \times 10^{11}$ cm from the jet base \citep{b23}.

  In models AD-1 and AD-2, we obtain the total mass-outflow rates 
  $\dot M_{\rm out}$ ($3.8 \times 10^{-5}$ and $1.3 \times 10^{-4} 
   {\rm M_\odot}$ yr$^{-1}$) and the axial mass-outflow rates
 $ \dot M_{1^\circ}$ ($9.2 \times 10^{-7}$ and $1.2 \times 10^{-6}  
 {\rm M_\odot}$ yr$^{-1}$) comparable with the observed wind ($\sim 10^{-5}$
  -- $10^{-4}  {\rm M_\odot}$ yr$^{-1}$) and jet
  ($\sim 10^{-7}$ -- $10^{-6} {\rm M_\odot}$ yr$^{-1}$) mass-ouflow rates,
  respectively.
 On the other hand, the mass-outflow rate $\dot M_{\rm out}$ in ML-1 and the
 axial outflow rate  $ \dot M_{1^\circ}$ in ML-2 are by more than 
 an order of magnitude lower compared with the corresponding observed 
 values of the wind and jets, respectively.

 The observed half opening angle of SS 433 jets is found to be very small, 
 about $\sim 1^{\circ}$, in the distant region of $\sim 10^{10}$ -- $10^{13}$ cm for the X-ray jets \citep{b23} and $10^{14}$ -- $10^{15}$ cm for optical
  jets \citep{b8}.
 Our computational domain ($r \sim 1.4 \times 10^{11}$ cm ) is located 
 roughly at the base of the observed X-ray jet, and the obtained
 half opening angle $\theta_{\rm c}$ for the high-velocity outflow with 
 radial velocities of $V_{\rm r} \ge 0.05c$ is 
 rather large, $\theta_{\rm c} \ge 30^{\circ}$.  
 The relatively large half opening angles of high-velocity flows, 
 $\sim 30^{\circ}$ -- 60$^{\circ}$, have been also found in other numerical
  studies of the supercritical accretion discs \citep{b6,b7,b28,b30,b27,b25},
  and in SS 433 such large opening angles of the high-velocity flow were at 
 odds with observations. The question in the previous studies was how to 
 collimate the high-velocity flow to a smaller angle. 
 In the present study, this is not a problem.
 Indeed, the mass-outflow rates obtained for the relativistic axial outflow
 in the half opening angle of $\sim 1^\circ$ are sufficiently high to 
 explain the observed mass-outflow rate of SS 433 jets, as far as the 
 results of AD-1 and AD-2 are considered.
 However, we have another question, why the relativistic axial outflow 
 in our simulations is rather undistingishable from the broader high-velocity
  outflow,
 while X-ray and optical jets of SS 433 are so brilliantly observed as 
 the hot relativistic streams with a small opening angle
 in a region far beyond the present computational domain.

  We interpret the small opening angles of the jets
 in terms of the proposal \citep{b23,b8} that the observed expansion velocity
 in the transverse direction of the jets  coincides with the sound velocity
 $c_{\rm s}$ in the region with a temperature $\sim 10^8$ K \citep{b23}
 and that the half opening angle $\theta_{\rm c}$ observed in the distant
 region
 at $r \gg 10^{11}$ cm should be equal to $\sim 2 c_{\rm s}/V_{\rm jet}$,
 where $V_{\rm jet} = 0.26c$.
 From our hydrodynamical results, in the axial flow  we have 
 $c_{\rm s} \sim 2.5\times 10^{-3}c$ and $T \sim 10^8$ K, which agrees very
 well with the idea above. 
 We suggest  that, if future observations resolve the jets deeper to
 the central source, the jets (high-velocity outflows) will be viewed
  with large opening angles, $\gg 1^\circ$. 

 Studying  H and He absorption lines during the precessing 
 motion of SS 433 disc, \citet{b8} derives the wind velocity $V_{\rm w}$ 
 of SS 433 as a function of the angle $\zeta$ from the disc plane:
  $V_{\rm w} \approx (8000\pm 100)\cdot 
 {\rm sin}^2 \zeta + 150 \;\; {\rm km\; s}^{-1}\;\; {\rm for} \;\; 
 0  \le \zeta \le 30^{\circ}$.
 This  gives velocities $V_{\rm w} = 400$ -- 2000 km s$^{\rm -1}$ 
 (0.001 -- 0.007$c$) for $10^{\circ} \le \zeta \le 30^{\circ}$, which
 are smaller by a factor of 2 -- 6 than the radial velocities
  $V_{\rm r}\sim$ 0.002 -- 0.04$c$ at the outer boundary in our simulations. 
  The high wind velocity $\sim$ 1500 km s$^{-1}$ (=0.005$c$) was suggested 
  from analysis of HeII emission near angles $\zeta \sim$ 70 -- 
  80$^{\circ}$ \citep{b8}.
  However, we obtain higher outflow velocities in these directions,
  0.08 -- 0.1$c$. Therefore, we have to presume
  some mechanism of decelerating the high-velocity flow in the distant
 region, such as interactions with the walls of gas cocoon surrounding the
 axial funnels where the jets expand.

 Although the absolute luminosity of SS 433 is not directly observed, 
 the kinematic luminosity $L_{\rm k}$ is generally estimated to be
  $\ge 10^{39}$ erg s$^{-1}$ \citep{b16}, which is consistent with
  $\sim$ 1 -- 2.5 $\times 10^{40}$ erg s$^{-1}$ obtained in all cases.
 The absolute luminosity of SS 433 
 is interesting because it is the maximum luminosity among
 the accreting stellar-mass black holes. The compact objects in the 
 recently observed ULXs are considered as stellar-mass 
 black hole or intermediate mass black hole, depending on whether their 
 luminosity far exceeds the Eddington luminosity $L_{\rm E}$ or lies at
 sub-Eddington values \citep{b24}. 
 The total luminosities $L$ obtained here are higher by a factor of 3
 than that given by the estimate $L/L_{\rm E}=0.6+0.7
  \;{\rm ln}\; \dot m_0$ obtained numerically in 1D model \citep{b20}
 and are higher by more than an order than the Eddington luminosity
 $1.5\times 10^{39}$ erg s$^{-1}$. 
 The collimated outflows in our simulations also lead to the outgoing 
 radiation directed along the rotational axis. 
 Actually, from the anisotropic radiation field obtained at the 
 outer boundary for each model, we have the radial component $F_{\rm r} 
  \sim 3 \times 10^{17}$ -- $5 \times 10^{18}$ erg s$^{-1}$ cm$^{-2}$ of
 the radiative flux $\bf F$ in the cone with a half opening angle 
 $30^{\circ}$.
 If an observer within this cone assume the radiative flux to be isotropic
 over the entire surface, the apparent luminosity is $7 \times 10^{40}$ -- 
 $10^{42}$ erg s$^{-1}$.  
 Thus, SS 433 can be a representative
 of the supercritically accreting stellar-mass black hole candidate,
 observed as the ULXs in nearby galaxies \citep{b14,b9,b2}.

 We should pay attention to the remarkable modulations of $\dot M_{\rm edge}$
 and $ L_{\rm d}$ in model AD-2.
 Their power spectra show QPOs-like signals at $\nu \sim 4 \times 
 10^{-2}$ -- $10^{-1}$ and 0.5 -- 2 Hz,  but only the
 larger amplitude modulations survive as small modulations of the 
 total luminosity with a quasi-period of $\sim$ 10 -- 25 s,
 as mentioned in subsection 4.2.
 These instabilities develop as the recurrent hot blobs with variable size
 rising through the accretion disc and the polar funnel region. 
 Such inhomogeneties could cause the luminosity modulations. 
 Although occasional massive jet ejections, which are recognized as a 
 clustering of flare events in radio light curves, were exhibited by 
 Microquasars, such as SS 433 \citep{b10} and GRS 1915+105 \citep{b11},
 X-ray observations of these events in SS 433 have been hardly performed 
 so far, because the massive jet ejections are rare, short, and aperiodic.
 Nevertheless, \citet{b17} reported recently a variety of new phenomena,
 including a QPO-like feature near 0.1 Hz, rapid time variability,
@and shot-like activities, and suggested that an irregular massive jet
 ejection might be caused by the formation of small plasma bullets or knots
 in the continuously emanating flow.
  We propose that the observed QPO-like phenomena in SS 433 can be 
 explained in terms of the recurrent hot blob phenomena found in model AD-2.

\section{Concluding remarks}
 
 We examined the jets and the disc of SS 433 with $\dot M_0
 \sim 600 \dot M_{\rm c}$ by time-dependent two-dimensional radiation
 hydrodynamical calculations, assuming $\alpha$-model for the viscosity. 
 The initial discs are given by 1D supercritical disc models with mass loss
 or advection. 
 As the result, the total luminosities obtained are 1 -- 2.5 $\times 10^{40}$
 erg s$^{-1}$, which are 6 to 10 times higher than the Eddington luminosity
 for $M=10{\rm M_{\odot}}$. 
 From the initial models with advection,
 we obtain the total mass-outflow rates $\dot M_{\rm out}\sim 4\times 
 10^{-5}$ and $10^{-4} {\rm M_{\odot}}$ yr$^{-1}$, and the relativistic
 axial outflow rates $\dot M_{1^\circ} \sim 10^{-6} {\rm M_{\odot}}$ 
 yr$^{-1}$. These outflow rates agree well with the observed mass-outflow
 rates of the wind and the jets in SS 433. 
 On the other hand, from the initial models with mass loss but without 
 advection, we obtain the total mass-outflow rates and the axial outflow
 rates smaller than or comparable to the observed rates of the wind and
 the jets respectively, depending on $\alpha$. 
 Still, while the mass-flow rate $\dot M_{1^{\circ}}$ of the axial
 outflow agrees well with the observed mass-outflow rate of SS 433 jets
 covering the same half opening angle $1^{\circ}$, a problem is to be
 solved, why in the simulations the axial outflow is not 
 distinguishable from the other high-velocity flow.

 The calculated radial velocities of the high-velocity outflow at the 
 outer boundary around the disc are larger by a factor of 2 -- 6 than
 the observed wind velocities \citep{b8}, which are given as a function
 of the elevation angle from the disc plane of SS 433.
 Thus, for the radial velocities to be consistent with the observed wind
 velocities, the outflow gas except the relativistic axial outflow must be 
 decelerated in the distant region beyond the present computational domain
 by some mechanism, such as interaction with the inhomogeneous matter or
 the gas cocoon around the SS 433 disc.

 The initial advective disc with large $\alpha$
 evolves to the gas-pressure dominant, optically thin state in the 
 inner region. The inner disc generates instabilities in agreement with the
 previous stability analyses of the advection-dominated optically thin discs. 
 As a result, we find remarkable modulations of the mass-inflow
 rate at the inner edge  and the disc luminosity. 
 The modulations of $\dot M_{\rm edge}$ present two types of variability;
 (1) the small amplitude variations with short time-scales of 0.5 -- 2 s, 
 (2) the large amplitude variations with long time-scales of $\sim$ 
 10 -- 25 s. 
 Only the variability (2) survives in the total luminosity because of the 
 atmospheric absorption around the accretion disc.
 The disc instability results in the recurrent hot blobs, which develop 
 outward and upward and produce QPOs-like phenomena in the total luminosity
 with the quasi-periods of $\sim$ 10 -- 25 s.
 The QPOs-like behavior of the luminosity and the hot blobs phenomena found
 here may explain the recent observations of a variety of new phenomena
 in SS 433, such as a QPO-like feature  near 0.1 Hz, rapid time variability,
  and a shot-like activity ascribed to the formation of small plasma bullets.
 
 \section*{Acknowledgments}

G. V. Lipunova has been supported by the Russian Foundation for Basic Research (project 09-02-00032). G. V. Lipunova is grateful to the Offene Ganztagesschule
of the Paul-Klee-Grundschule (Bonn, Germany) and Stadt Bonn for providing a
 possibility for her full-day scientific activity.

\label{lastpage}

\end{document}